%% file: ms.tex
\documentclass[sn-apa]{sn-jnl}\usepackage[]{graphicx}\usepackage[]{xcolor}
\makeatletter
\def\maxwidth{ %
  \ifdim\Gin@nat@width>\linewidth
    \linewidth
  \else
    \Gin@nat@width
  \fi
}
\makeatother

\definecolor{fgcolor}{rgb}{0.345, 0.345, 0.345}

\usepackage{framed}
\makeatletter
 {\par\unskip\endMakeFramed%
 \at@end@of@kframe}
\makeatother

\definecolor{shadecolor}{rgb}{.97, .97, .97}
\definecolor{messagecolor}{rgb}{0, 0, 0}
\definecolor{warningcolor}{rgb}{1, 0, 1}
\definecolor{errorcolor}{rgb}{1, 0, 0}
\newenvironment{knitrout}{}{} 

\usepackage{alltt}

\newcommand{\Sconcordance}[1]{%
  \ifx\pdfoutput\undefined%
  \csname newcount\endcsname\pdfoutput\fi%
  \ifcase\pdfoutput\special{#1}%
  \else%
   \begingroup%
     \pdfcompresslevel=0%
     \immediate\pdfobj stream{#1}%
     \pdfcatalog{/SweaveConcordance \the\pdflastobj\space 0 R}%
   \endgroup%
  \fi}
\input{ms-concordance}

\newcommand{\Sum}{\sum\limits}

\usepackage{amsmath}
\usepackage{url}

\jyear{2022}%

\theoremstyle{thmstyleone}%
\theoremstyle{thmstyletwo}%

\theoremstyle{thmstylethree}%

\IfFileExists{upquote.sty}{\usepackage{upquote}}{}
\begin{document}

\title[Data Sharpening in AR(1) Models]{On Data Sharpening in Nonparametric Autoregressive Models}

\author[1]{\fnm{Simon} \sur{Snyman}}\email{simon10snyman@gmail.com}

\author*[1]{\fnm{Lengyi} \sur{Han}}\email{lengyi.han@ubc.ca}

\author[1]{\fnm{W. John} \sur{Braun}}\email{john.braun@ubc.ca}

\affil*[1]{\orgdiv{I.K. Barber Faculty of Science}, \orgname{UBC}, \orgaddress{\street{1177 Research Road}, \city{Kelowna}, \postcode{V1V 1V7}, \state{BC}, \country{Canada}}}


\abstract{Data sharpening has been shown to reduce bias in nonparametric regression
  and density estimation.  Its performance on nonlinear
  first order autoregressive models is studied theoretically and numerically in this paper.   Although
  the asymptotic properties of data sharpening are not as favourable in the presence
  of serial dependence as in 
  bivariate regression with independent responses, it is still found to reduce bias under mild conditions on the autoregression
  function.  Numerical comparisons with the bias reduction method of \citet{cheng2018bias} 
  indicate that data sharpening is competitive in this setting.
 }


\maketitle

\section{Introduction}\label{sec1}
We consider data in the form of a stationary time series $z_1, z_2, \ldots, z_n$.  
The model we will study is 
\begin{equation}\label{eq:nonlinearAR1} z_t = g(z_{t-1}) + \sigma(z_{t-1}) \varepsilon_t \end{equation}
where $g(\cdot)$ is a function with at least two continuous derivatives at an evaluation point $z$, and $\{\varepsilon_t\}$ is
a sequence of independent noise terms with  mean $E[\varepsilon_t] = 0$ and 
variance  $\mbox{Var}(\varepsilon_t ) =  1$.  Also, $\varepsilon_t$ is independent of any member
of the sequence  $z_{t-1}, z_{t-2}, \ldots$.  
This model is a special case of the nonparametric autoregressions considered by
\citet{fan2008nonlinear} where local polynomial regression and spline smoothing are proposed to estimate
$g(\cdot)$.  
In the current paper, we consider two of these estimators: the local constant regression estimator for $g(z)$ of the form 
    \[\widehat{g}(z) =  \frac{\Sum_{i = 2}^{n} K_{h} (z_{i-1}-z) z_{i}}{\Sum_{i = 2}^{n} K_{h} (z_{i-1}-z)} \] 
and the local linear regression estimator
    \[\widehat{g}(z) =  \frac{\left(\Sum_{i = 2}^{n} K_i z_{i}\right)\left(
    \Sum_{i = 2}^{n}(z_{i-1} - z)^2 K_i
     \right)
     - \left(\Sum_{i = 2}^{n} (z_{i-1} - z)K_i\right)\left(
    \Sum_{i = 2}^{n} (z_{i-1} - z)K_i z_i\right)
     }{\left(\Sum_{i = 2}^{n} K_i \right) \left( \Sum_{i=2}^n (z_{i-1}-z)^2 K_i \right) - \left(\Sum _{i=2}^n (z_{i-1}-z)K_i\right)^2}\] 
     where $K_i = K_h(z_{i-1} - z)$.   Here, the kernel $K_h(.)$ is assumed to be a symmetric probability density
     function with scale parameter $h$ (usually called the bandwidth).  
Both of these estimators are linear in the responses, so each is representable in the general form
\begin{equation}\label{eq:linear}\widehat{g}(z) = \sum_{i=2}^n A_h(z_{i-1} - z) z_i \end{equation}
where, for example, $A_h(z_{i-1} - z) = \frac{1}{\sum_{j=2}^n K_j} K_i$ in the local constant case.  
For additional information on kernel smoothing, we refer to  \citet{fan2008nonlinear} for
     their excellent summary of  the salient
     aspects of local regression.

The model (\ref{eq:nonlinearAR1}) can also be viewed as a special case of the functional autoregressive models first studied by \citet{chen1993functional} where we might write $g(z) = z m(z)$ for some function $m(\cdot)$.  A
sufficient condition for such a process to be stationary is $\vert m(z) \vert \leq c < 1$ for all $z$.  
\citet{park2015varying} provide a review of both kernel-based and spline-based methods used in the fitting of 
varying coefficient regression methods and briefly consider the autoregressive model of \citet{chen1993functional} as
a special case.  \citet{cai2009functional} consider nonlinear nonstationary time series from the standpoint of a
varying coefficient model, and they employ a two-step nonparametric kernel smoothing procedure analogous
to profile likelihood methods with
local linear regression to handle nonstationary covariates.  They are able to achieve an optimal pointwise convergence
rate for their method under fairly mild assumptions, and they are able to prove asymptotic normality under stronger assumptions.  

\citet{geng2022estimation} considers functional coefficient AR models with measurement error, and introduces a
local linear regression procedure that is bias-corrected through explicit estimation of the leading bias term 
based on kernel density estimation.  Knowledge of the measurement error standard deviation is usually required for the
method to work.   We also note the paper by \citet{lin2019global} which describes a very general autoregressive model as
well as a considerable number of references to research conducted  on these kinds of model over the past thirty years.  

\section{Bias and Bias Reduction Procedures}

Bias reduction in local regression has been considered by several authors.  
 \citet{choi2000data} proposed a number of forms of
data sharpening to reduce bias in local polynomial regression.  A simple version involved constructing the
nonparametric estimate, adding
the residual to the response data and re-fitting the nonparametric regression using the sharpened responses.  The
technique was found to be highly effective in reducing bias for
local constant and local linear regression.  
\citet{cheng2018bias} proposed a regression-based method for bias reduction in local linear
regression as well as improved variance estimation in nonparametric regression.   
We should also mention
the work of \citet{he2019} which concerns data sharpening for regression models with dependent errors.  The
model considered there is quite different from the one we consider in this paper, however.  

We now consider the situation for autoregression. 
We begin with the assumption that the autoregression function $g(\cdot)$ has two continuous derivatives at the evaluation point $z$ and $h = O(n^{1/3})$ (local constant case) or $h = O(n^{1/5})$ (local linear case), and assume the kernel $K(z)$ has bounded support.   Assume also that $\sigma(\cdot)$ and the stationary distribution 
$f(\cdot)$ are continuous are continuous at $z$.  
Section 6.6.2 of \citet{fan2008nonlinear} summarizes additional conditions on the time series which are sufficient for local constant and local linear autoregression estimators to have bias of order $O(h)$ and $O(h^2)$, respectively, when the evaluation point is in the interior
of the range of the data.  The conditions given there are also sufficient to  yield central limit
theorems for these estimators. 
Our present purpose is to determine whether bias reduction strategies that are known to work well in local constant and
local linear regression continue to perform well in the autoregression setting.  

One of the data sharpening strategies of \citet{choi2000data} is to add residuals to the response values to obtain sharpened
responses.  Adapting this approach to the autogressive model, we propose the following data sharpening procedure:
\begin{enumerate}
\item Set $z^\star_t = 2z_t - \widehat{g}(z_{t-1})$. 
\item The sharpened local constant and local linear autoregression estimators are then
\[ \widehat{g}^\star(z) =  \sum_{i=2}^n A_h(z_{i-1} - z) z^\star_i \]
where the precise form of $A_h(\cdot)$, as given in (\ref{eq:linear}),  is dictated by which type of estimator is used. 
\end{enumerate}
Note that the sharpened data only appear as ``responses'', and the original data continue to be
used as ``design points''.   Data sharpening for local constant and local linear regression has been shown to 
successfully decrease the order of the asymptotic bias.  We will find both theoretically and numerically that
the situation is a little different for local autoregression.

The \citet{cheng2018bias} bias reduction methods can also be readily adapted to the local linear autoregression problem, because they are
designed to explicitly remove the $O(h^2)$ term in the asymptotic bias.  We consider one of the methods for the
purpose of comparing with the data sharpening procedure.  The method is as follows:
\begin{enumerate}
\item Calculate $\widehat{g}(z; h)$ using a sequence of bandwidths $h_1, h_2, \ldots, h_m$ (we use the same choices
as in \citet{cheng2018bias}); set $Y_h = \widehat{g}_(z; h)$, for $h \in \{h_1, h_2, \ldots, h_m\}$. 
\item Estimate $\beta_0$ and $\beta_1$ in the simple regression model
\[ Y_h = \beta_0 + \beta_1 h^2 + \epsilon_h. \]
\item Define $\tilde{g}(z) = \widehat{\beta}_0$ to be the bias-reduced estimator for $g(z)$. 
\end{enumerate}
Note that this method is virtually identical to the proposal given by \citet{cheng2018bias}.  The only difference is
the form of the input data.  

\subsection{A Decomposition of the Sharpening Bias}

Bias reduction is possible with data sharpening but is not guaranteed in small samples, as we indicate here.
The sharpening transformation can be written as 
\[ z_i^\star  = z_i + g(z_{i-1}) + \varepsilon_i \sigma(z_{i-1}) - \widehat{g}(z_{i-1})\]
where either the local constant or local linear estimator could be in use.  
Using the notation introduced at (\ref{eq:linear}), we can further write
\[ z_i^\star = z_i + g(z_{i-1}) + \varepsilon_i \sigma(z_{i-1}) - \Sum_{t=2}^n A_h(z_{t-1} - z_{i-1}) g(z_{t-1})
- \Sum_{t=2}^n A_h(z_{t-1} - z_{i-1}) \varepsilon_t \sigma(z_{t-1}). \]
Thus, 
\begin{eqnarray} \widehat{g}^\star(z) = \widehat{g}(z) + \Sum_{i=2}^n A_h(z_{i-1} - z) g(z_{i-1}) - \Sum_{i=2}^n A_h(z_{i-1} - z) \Sum_{t=2}^n A_h(z_{t-1} - z_{i-1}) g(z_{t-1}) \nonumber \\
+ \Sum_{i=2}^n A_h(z_{i-1} - z) \varepsilon_i \sigma(z_{i-1}) - \Sum_{i=2}^n A_h(z_{i-1} - z) \Sum_{t=2}^n A_h(z_{t-1} - z_{i-1}) \varepsilon_t \sigma (z_{t-1})
. \nonumber \end{eqnarray}
Taking expectations and collecting terms, we see that the bias for a data sharpened local regression estimator can be 
decomposed as 
\begin{equation}\label{eq:decomposition} B^\star(z) = B_g(z) - B_{\widehat{g}}(z) - \mbox{Err}(z). \end{equation}
Here $B_g(z) = E[\widehat{g}(z) - g(z)]$ is the bias in the unsharpened local regression estimator, 
\[B_{\widehat{g}}(z) = E\left[\Sum_{i=2}^n A_h(z_{i-1} - z) \left( \Sum_{t=2}^n A_h(z_{t-1} - z_{i-1}) g(z_{t-1}) 
- g(z_{i-1})\right)\right], \]
and 
\[ \mbox{Err}(z) = E\left[\Sum_{i=2}^n A_h(z_{i-1} - z) \left(\Sum_{t=2}^n A_h(z_{t-1} - z_{i-1}) \varepsilon_t \sigma (z_{t-1})
- \varepsilon_i \sigma(z_{i-1})\right)\right]. \] 
 The last two terms in (\ref{eq:decomposition}) warrant further comment.  The $B_g(z)$ term is similar to the usual
 correction induced by the data sharpening procedure in local regression. This term decreases the
 asymptotic order represented in the $B_g(z)$ term in both the local regression and local autoregression situations.
 The $\mbox{Err}(z)$ term arises only in the autoregression scenario and causes a somewhat unpredictable level
 of distortion.  In local regression, the fact that the condition mean of $\varepsilon_i$ is 0 is sufficient to cause
 this term to vanish.  In local constant autoregression, the dependence structure prohibits this from happening, as
 pointed out by \cite{fan2008nonlinear}.  These authors also point out that the dependence structure is substantially
 weakened by the presence of a kernel, so that they are able to use central limit theorem arguments to develop
 bias expressions instead of a direct approach involving explicit calculation of expectations.  
 
\subsection{Numerical Evidence}
 To understand the impact of $\mbox{Err}(z)$ on the bias, we studied its behaviour by simulation.  We use 
 the function \verb!sharpARlocpoly()! in the \textit{RCMinification} R package \citep{Han2023} for this purpose. For each 
 simulation scenario, 500 samples of size 50, 100, and 200 were taken from the model (\ref{eq:nonlinearAR1}) 
 using normally distributed noise with $\sigma = 0.5$.  For each sample, local constant and local linear autoregression curves were estimated, 
 together with their data sharpened counterparts.  For this study, bandwidths were fixed for each sample size: $h = 0.3$
 when $n = 50$, $h = 0.25$ when $n = 100$ and $h = 0.2$ when $n = 200$. 
 
 Bias was estimated for the two estimation procedures as a function of
 the evaluation points which were taken in the interval $[-1, 1]$.  Estimates of $B_{\widehat{g}}(z)$ and $\mbox{Err}(z)$  
 were also computed for $z \in [-1, 1]$.     We considered
 two test autoregression functions for this study:  $g(x) = x \sin(x)$ and $g(x) = \cos(x)$.  The first function
 corresponds to a stationary time series and the latter gives rise to a nonstationary time series.   Figure \ref{fig:biascomponents}
displays the results for the first function and Figure \ref{fig:biascomponents2} shows the results for the second function.

In each figure, the top three panels display results for local constant autoregression estimation and the bottom three panels
contain the local linear autoregression results.  Left panel sample sizes are $n = 50$, middle panel samples are $n = 100$,
and right panel sample sizes are $n = 200$.  Each panel contains five curves.  The solid black curve is the bias of the local
autoregression estimator.    The solid blue curve is the bias of the sharpened autoregression estimator.  The short dashed
blue curve represents $B_{\widehat{g}}(z)$, the dotted red curve represents $\mbox{Err}(z)$, and the long-dashed green
curve represents the combination of the three components:   $B_g(z) - B_{\widehat{g}}(z) - \mbox{Err}(z).$

The two figures tell somewhat similar stories.  As would be expected, the magnitude of the bias is generally smaller when local linear
estimation is used, particularly at the boundaries.  Improvements in both estimators are seen as the sample size increases.  The data sharpening procedure substantially decreases bias at the boundaries in the local constant cases, and less so in the local linear cases
where the boundary effect is less pronounced - similar to what has been reported in local linear regression by many researchers starting with Fan and his colleagues.  

The behaviour of $\mbox{Err}(z)$ appears to improve as the sample size increases, although its effects in local linear autoregression
appear to be somewhat more impactful than in local constant autoregression.  The combination  $B_g(z) - B_{\widehat{g}}(z) - \mbox{Err}(z)$  does not match the bias in data sharpening very closely when $n = 50$,  The differences are most striking in the local
linear cases.  Even when $n = 200$, there is fairly serious distortion caused by $\mbox{Err}(z)$ to the bias in data sharpened local linear
autoregression  near the left boundary for the $\cos(.)$ target function.   The rest of the function appears to have been estimated
with very little bias when data sharpening has been applied.   

Our conclusion from this brief study is that gains can be made through the use of data sharpening, particularly with local
linear autoregression, but the performance will often be inferior to what would be expected from applying data sharpening
to local linear regression.

\begin{knitrout}
\definecolor{shadecolor}{rgb}{0.969, 0.969, 0.969}\color{fgcolor}\begin{figure}
\includegraphics[width=\maxwidth]{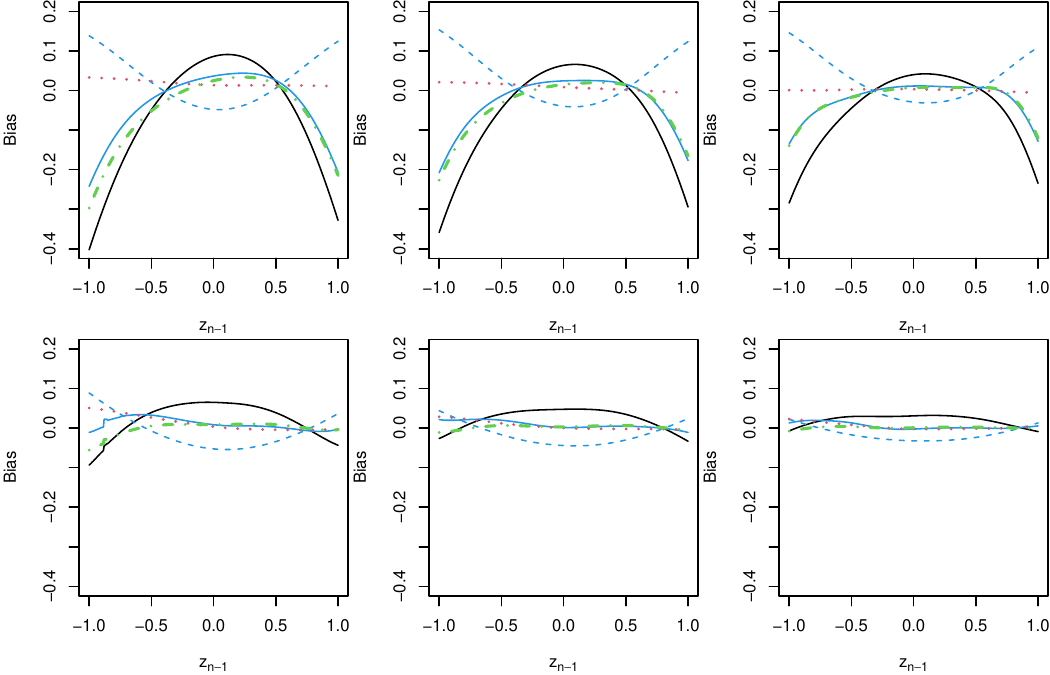} \caption[Bias of the data sharpened local autoregression estimator and its relation to the bias of the unsharpened local estimator and the noise component]{Bias of the data sharpened local autoregression estimator and its relation to the bias of the unsharpened local estimator and the noise component. The underlying autoregression function is $g(x) = x\sin(x)$. The solid black curve represents bias for the local estimator, the blue curve corresponds to the data sharpened estimator, the red dotted curve represents the effect of the noise, and the short-dashed blue curve represents the effect of sharpening on the bias after removing the noise effect. The top panels contain local constant results, and the bottom panels contain local linear results.  Sample sizes increase from the left panel to the right panel:  $n = $ 50, 100 and 200.  }\label{fig:biascomponents}
\end{figure}

\end{knitrout}

\begin{knitrout}
\definecolor{shadecolor}{rgb}{0.969, 0.969, 0.969}\color{fgcolor}\begin{figure}
\includegraphics[width=\maxwidth]{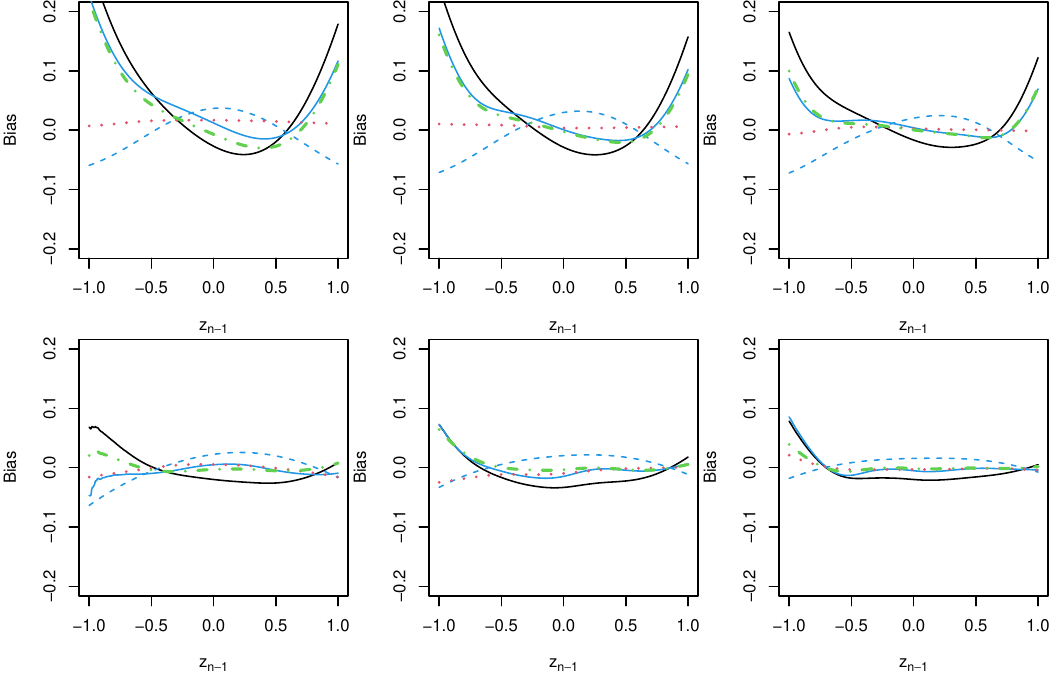} \caption{A demonstration of the bias decomposition for estimates of the function $g(x) = \cos(x)$. See caption to Figure \ref{fig:biascomponents} for legend details.}\label{fig:biascomponents2}
\end{figure}

\end{knitrout}

\subsection{Cheng's Method Versus Data Sharpening}

We now compare Cheng's bias reduction method with data sharpening in the case of local linear autoregression via a
simulation study.  Unlike the study in the previous section where we aimed to understand the data sharpening
mechanism and used the same bandwidth for all simulation experiments, this time we need to use a data-driven
bandwidth in order to emulate the practical circumstances that would face the data analyst when making the
choice between the two methods.  

For each 
 simulation scenario, 5000 samples were taken from the model (\ref{eq:nonlinearAR1}) 
 using normally distributed noise with $\sigma = 0.5$.  For each sample, local  linear autoregression curves were estimated, 
 together with their data sharpened and Cheng's method counterparts.  The bandwidth $h$ is calculated 
 for the local linear regression method using the \verb!thumbBw()! function that is
available  in the \textit{locpol} package \citep{locpol}.   This is adjusted for data sharpening through multiplication
by $n^{4/45}$ (in order to take advantage of any potentially reduction in bias).  The rule-of-thumb bandwidth was also used to construct the sequence of bandwidths for use 
in Cheng's method: $h_j = (1+(j-1)/10)h, j = 1, 2, \ldots, 11$.  
  
 Bias was estimated for the three estimation procedures as a function of
 the evaluation points which were taken to be equally spaced in the interval $[-0.5, 0.5]$.   
 The average absolute error was also calculated at the same points.  
 We considered
 four test autoregression functions for this study: $g(x) = \cos(x)$, $g(x) = x\cos(x)$,  $g(x) = \sin(x)$,  and $g(x) = x \sin(x)$.
 The first process is nonstationary while the last three are stationary.  
  Figure \ref{fig:biascomparison50}
graphically displays the results for sample sizes of $n = 50$, where each row of the layout corresponds to one of the
four functions. The absolute bias function estimates are in the left panels and the mean absolute error function estimates
are in the right panels, where the solid black curve represents raw local linear estimation, the dashed blue curve represents
sharpened estimation, and the dotted red curve represents Cheng's method.   Figures \ref{fig:biascomparison100}  and
\ref{fig:biascomparison200} display similar information but for sample sizes $n = 100$ and $n = 200$, respectively.

\begin{knitrout}
\definecolor{shadecolor}{rgb}{0.969, 0.969, 0.969}\color{fgcolor}\begin{figure}
\includegraphics[width=\maxwidth]{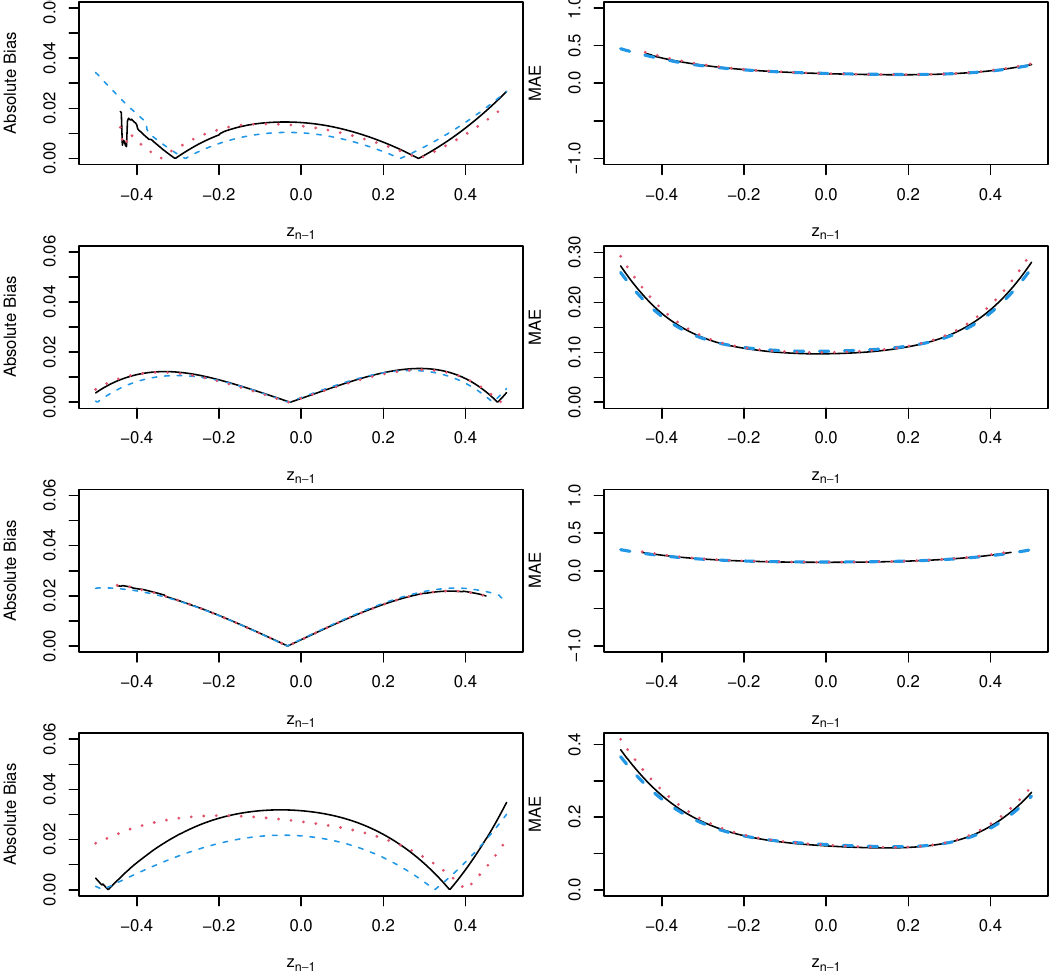} \caption[Bias (left panels) and Mean Absolute Error (right panels) comparisons]{Bias (left panels) and Mean Absolute Error (right panels) comparisons: local linear autoregression (solid black curve), data sharpening (dashed blue curve) and Cheng's bias reduction method (dotted red curve). Target functions are from top to bottom: $\cos(x), x\cos(x), \sin(x)$ and $x\sin(x)$.  Sample size is 50.}\label{fig:biascomparison50}
\end{figure}

\end{knitrout}

\begin{knitrout}
\definecolor{shadecolor}{rgb}{0.969, 0.969, 0.969}\color{fgcolor}\begin{figure}
\includegraphics[width=\maxwidth]{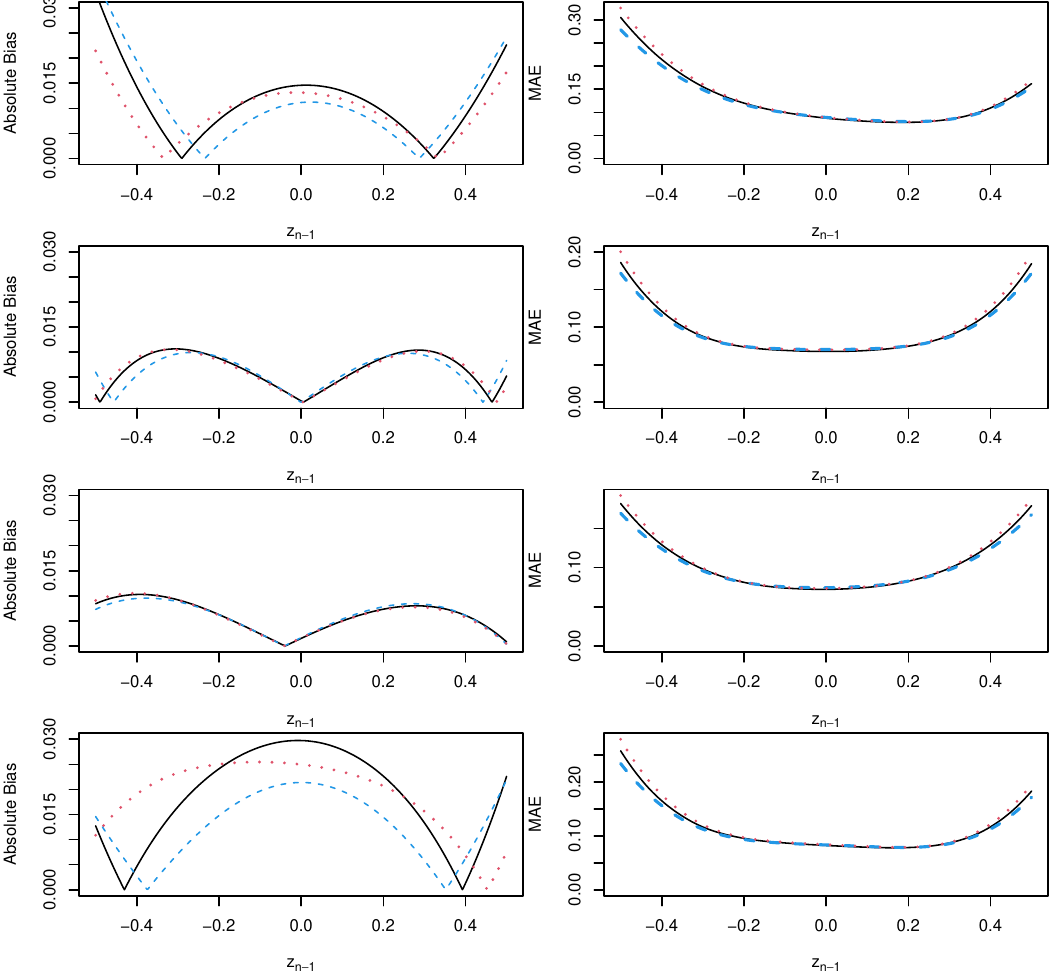} \caption{Bias comparisons: local linear autoregression, data sharpening and Cheng's bias reduction method. Sample size is 100. See Figure \ref{fig:biascomparison50} caption for more legend information.}\label{fig:biascomparison100}
\end{figure}

\end{knitrout}

\begin{knitrout}
\definecolor{shadecolor}{rgb}{0.969, 0.969, 0.969}\color{fgcolor}\begin{figure}
\includegraphics[width=\maxwidth]{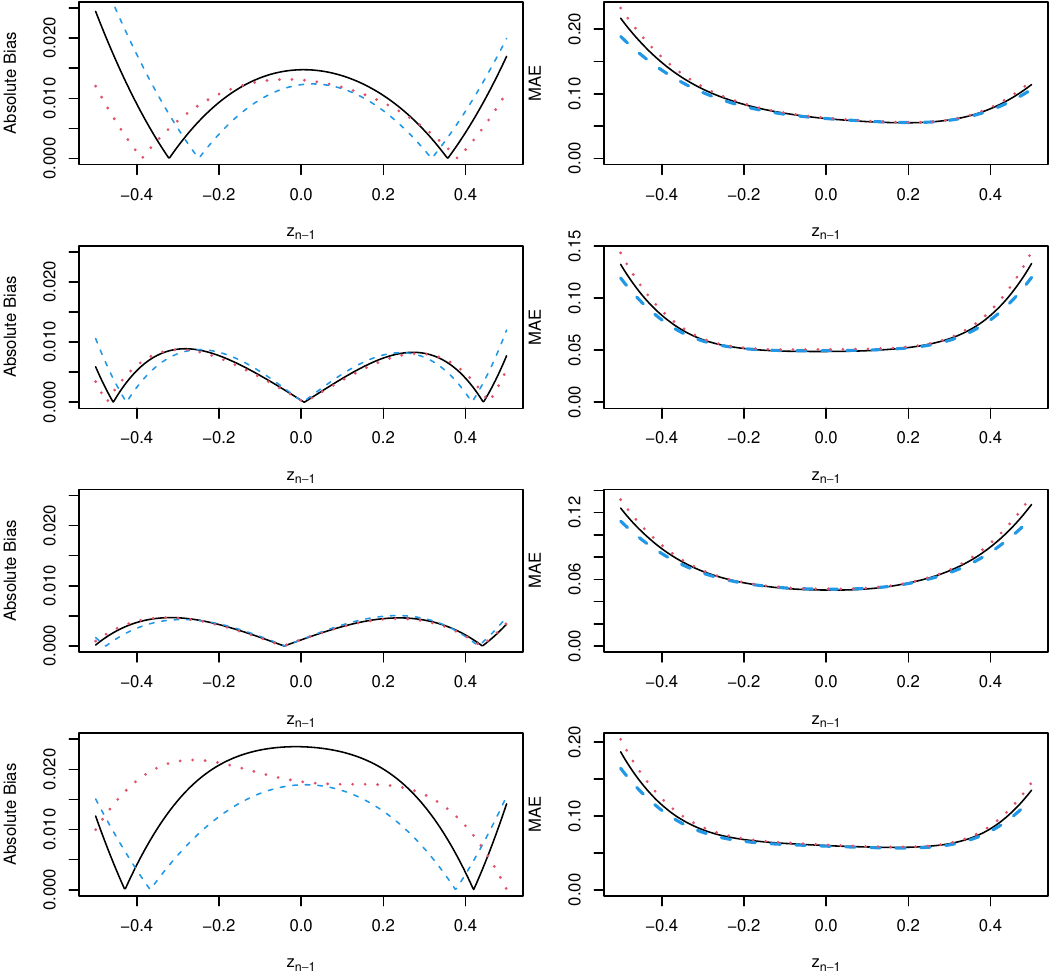} \caption{Bias comparisons: local linear autoregression, data sharpening and Cheng's bias reduction method. Sample size is 200.   See Figure \ref{fig:biascomparison50} caption for more legend information.}\label{fig:biascomparison200}
\end{figure}

\end{knitrout}

The results suggest that when the sample size is small ($n = 50$), bias is similar among the three methods, though data
sharpening appears to provide some improvement in the case of $g(x) = x\sin(x)$.  When $n = 100$, data sharpening
continues to outperform the other methods for this function, and it does better in the interior
of the range in the case of $g(x) = \cos(x)$.  In that case, Cheng's method performs somewhat better than the
other two methods near the boundaries of the range.    The Mean Absolute Error plots show that the three
methods all perform similarly, with data sharpening often very slightly superior to the others.   Thus, there
appears to be little harm in performing data sharpening, remembering that its reduced bias allows for easier interpretation
in practical situations.  

\section{Applications to Data Analysis}

We consider two datasets here to illustrate the use of the local linear autoregression estimator
and its bias-reduced forms.  

\subsection{Earthquake Data}

This dataset represents the annual numbers of earthquakes with a higher magnitude than 7 as measured on the Richter Scale between the years 1916 and 2015 (It is available from \url{https://online.stat.psu.edu/stat501/book/export/html/995}).   A plot
of the estimated autocorrelation function for this dataset indicates that there is slight autocorrelation at lags 1 and 3. Since
the partial autocorrelation function suggests that only the lag 1 autocorrelation be taken seriously, we 
tentatively put forward the model
\[ z_t = g(z_{t-1}) + \varepsilon_t\]
for these data.  
Applying the local linear autoregression estimator, we obtain the solid black curve in the left panel of Figure \ref{fig:earth}
as our estimate of $g(z)$.

There is a suggestion of nonlinearity in the curve, so in 
the interest of testing whether this is inconsistent with a linear model, we constructed a bootstrap test as follows.
First, we estimated the parameters for a first order autoregressive process, obtaining the fitted model
\[ z_t = 0.2692 z_{t-1} + \varepsilon_t + 12.59 \]
where the variance of $\varepsilon_t$ was estimated to be 16.56. Under a hypothesis that this is the true model,
we would have  $g(z) = 0.2692z + 12.59$.  
Simulating from this model 500 times yielded 500 time series to which we applied the local linear autoregression
estimator, yielding 500 estimates of $g(z)$ at 401 evaluation points in the range of data.  Pointwise upper and lower test limits were obtained by finding the 0.025 and 0.975 quantiles of the $g(z)$ estimates for each evaluation point $z$. 
The intervals between these test limits are shaded in the figure with the interpretation that if the 
original curve falls outside of these limits, there is evidence of true nonlinearity.  Since the curve easily
falls within the limits, we can make no such conclusion.  

The middle panel and right panel contain the analogous data sharpened and Cheng's method results. Neither
case provides a contradiction to the original result.

\begin{knitrout}
\definecolor{shadecolor}{rgb}{0.969, 0.969, 0.969}\color{fgcolor}\begin{figure}
\includegraphics[width=\maxwidth]{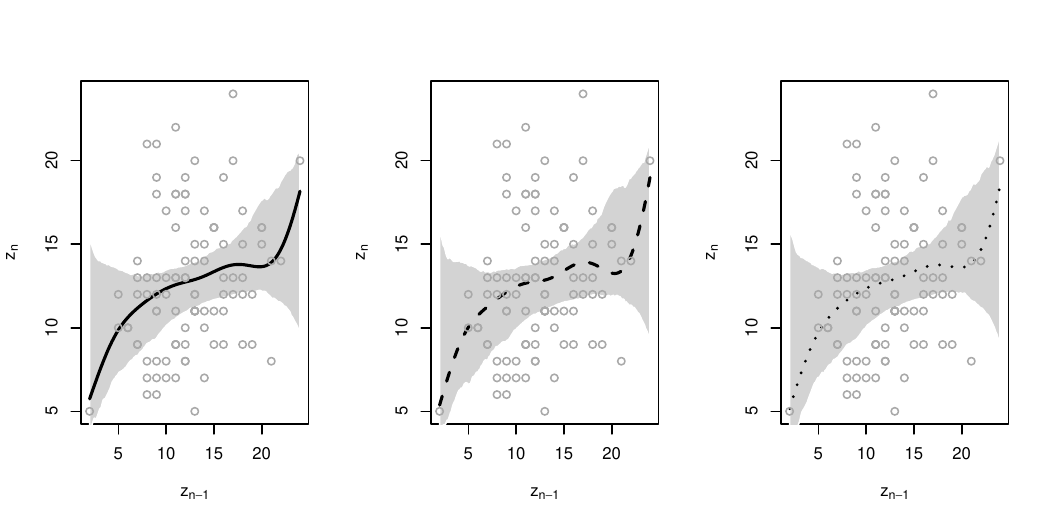} \caption[Estimates of first order nonlinear autoregressive model for the earthquakes data]{Estimates of first order nonlinear autoregressive model for the earthquakes data.  Grey areas correspond to pointwise 95\% bootstrap test bands.  Left panel: local linear estimate; center panel: data sharpened local linear estimate; right panel: Cheng's bias-reduced local linear estimator.}\label{fig:earth}
\end{figure}

\end{knitrout}

\subsection{Application to lynx AR(2) residuals}

The lynx data are available in R \citep{R}, and they were analyzed by \citet{fan2008nonlinear} using
the local linear autoregression procedure.  Here, we take a somewhat different tack.   Instead
of using the autoregression estimator to forecasting as in \citet{fan2008nonlinear}, we illustrate
how it can be used for model criticism.  In this case, a second order autoregressive model seems to 
give a reasonable fit to the square root values of the annual counts.   Our approach then will be
similar to our testing approach for the earthquake dataset, except in this case, we will perform
the test on the residuals after fitting the second order autoregressive model.   Specifically,
we will check if there is first lag dependence in the residuals that is not captured by the model.
The fitted model is 
\[ z_t = 1.3088 z_{t-1} - 0.7104 z_{t-2} + 31.128 + \varepsilon_t \]
where $\varepsilon_t$ has variance 76.51. 
Interestingly, the first lag autocorrelation in the residuals from this model is close to 0.   However,
if we view the residual process as
\[ e_t = g(e_{t-1}) + \epsilon_t \]
for some unknown function $g(.)$, we can apply local linear autoregression, and the bias reduction methods
to estimate $g$.    The local linear, data sharpened and Cheng's method estimates are plotted in the
three panels of Figure \ref{fig:lynx}.    The bootstrap test bands are calculated in almost the same
way as for the earthquake data analysis.  That is, we simulated from the fitted AR(2) model, and
estimated the AR(2) parameters from the simulated data.    Residuals were computed, and curve estimates of $g$
were obtained by applying all three estimation methods to the simulated residuals, knowing that the true function is $g(z) \equiv 0$.  Simulations were conducted 500 times, and
2.5 and 97.5 percentiles were obtained at each of 401 evaluation points.  These were used to construct lower and upper test
bands for each of the three methods.  The results are plotted as the shaded regions in Figure \ref{fig:lynx}.  
 
\begin{knitrout}
\definecolor{shadecolor}{rgb}{0.969, 0.969, 0.969}\color{fgcolor}\begin{figure}
\includegraphics[width=\maxwidth]{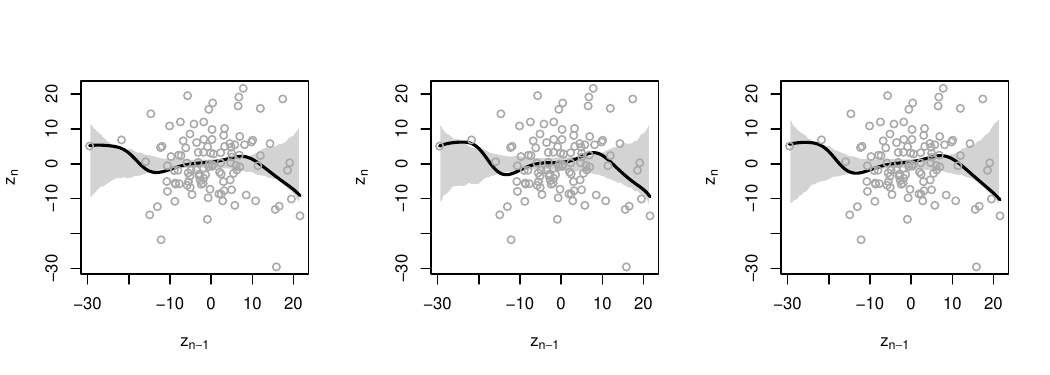} \caption[Estimates of first order nonlinear autoregressive model for residuals after fitting a second order linear autoregressive model to the lynx data, on the square root scale]{Estimates of first order nonlinear autoregressive model for residuals after fitting a second order linear autoregressive model to the lynx data, on the square root scale.  Grey areas correspond to pointwise 95\% bootstrap test bands.  Left panel: local linear estimate; center panel: data sharpened local linear estimate; right panel: Cheng's bias-reduced local linear estimator.}\label{fig:lynx}
\end{figure}

\end{knitrout}

This time, the data sharpened curve falls outside of the test bands in three different locations.  Results from Cheng's method
and the local linear method are more borderline.    Overall, we have some evidence from these results to suggest
that the linear AR(2) model is not completely satisfactory for these data.  

\section{Conclusion}\label{sec13}

In this paper, we studied data sharpening for local constant and local linear autoregression modelling
of time series.  We found that, unlike data sharpening for local regression, the amount of bias reduction
achieved is not as predictable, though in many situations, a reduction can be attained.   Overall absolute
error is often marginally decreased with data sharpening.    

Although the method of \citet{cheng2018bias} may have a clearer theoretical justification, its performance
in bias reduction is often inferior to that of data sharpening.  

\bmhead{Acknowledgments}

This work has been supported by a Discovery Grant from the Natural Sciences and 
Engineering Research Council of Canada (NSERC).    SS also acknowledges additional
support from the NSERC USRA program.




\end{document}

%% file: ms-concordance.tex
\Sconcordance{concordance:ms.tex:ms.Rnw:%
1 27 1 50 0 3 1 10 0 240 1 1 3 3 1 3 0 4 1 3 0 31 1 1 4 8 1 1 3 7 1 1 3 %
7 1 1 3 44 1 3 0 29 1 3 0 43 1}